\documentclass[prb,preprint,showpacs,preprintnumbers,amsmath]{revtex4}
\usepackage{graphicx}
\usepackage{dcolumn}
\usepackage{bm}

\begin{document}
\title{Kinetic stabilization of Fe film on GaAs(100):
An {\it in situ} x-ray reflectivity Study}

\author{T. C. Kim}
\affiliation{Department of Materials Science and Engineering,
Gwangju Institute of Science and Technology, Gwangju 500-712,
Korea}

\author{J.-M. Lee}
\affiliation{School of Physics \& Center for Strongly Correlated
Materials Research, Seoul National University, Seoul 151-742,
Korea}

\author{Y. Kim}
\affiliation{Department of Materials Science and Engineering,
Gwangju Institute of Science and Technology, Gwangju 500-712,
Korea}

\author{D. Y. Noh}
\affiliation{Department of Materials Science and Engineering,
Gwangju Institute of Science and Technology, Gwangju 500-712,
Korea}

\author{S.-J. Oh}
\affiliation{School of Physics \& Center for Strongly Correlated
Materials Research, Seoul National University, Seoul 151-742,
Korea}

\author{J.-S. Kim}
\affiliation{Department of Physics,
Sook-Myung Women's University, Seoul 140-742, Korea}
\date{\today }

\begin{abstract}

We study the growth of the Fe films on GaAs(100) at a low
temperature, 140 K, by $in$-$situ$ UHV x-ray reflectivity using
synchrotron radiation. We find rough surface with the growth
exponent, $\beta_S$ = 0.51$\pm$0.04. This indicates that the growth
of the Fe film proceeds via the restrictive relaxation due to
insufficient thermal diffusion of the adatoms. The XRR curves are
nicely fit by a model with a uniform Fe film, implying that the
surface segregation and interface alloying of both Ga and As are
negligible. When the Fe film is annealed to 300 K, however, the
corresponding XRR can be fit only after including an additional
layer of 9 $\AA$ thickness between the Fe film and the substrate,
indicating the formation of ultrathin alloy near the interface. The
confinement of the alloy near the interface derives from the fact
that the diffusion of Ga and As from the substrate  should proceed
via the inefficient bulk diffusion, and hence the overlying Fe film
is kinetically stabilized.

\end{abstract}
\pacs{68. 55. -a, 68. 35. Fx, 68. 35. Ct}
\maketitle

Recently,  the ferromagnetic metal-semiconductor heterostructures
has drawn intense attention for their application
to the spintronic devices such as spin-polarized
field effect transistors and spin-polarized light emitting diodes.~\cite{Wolf}
Such spintronic devices require
highly polarized spin source and efficient spin injection through
the interface of the heterostructures. The Fe/GaAs system has been
studied as a prototype of the heterostructures, because high
quality, epitaxial bcc Fe films can be grown  on GaAs surfaces due
to their small lattice mismatch only by  $\sim$ 1.3
$\%$~\cite{Waldrop,Zolfl}. Moreover, the spin injection from the
Fe film into GaAs(100) has been realized even at room
temperature~\cite{Zhu}. However, the system still has enjoyed
limited success, because the magnetic polarization of the Fe film
and in turn spin injection efficiency has suffered from serious
degradation due to the formation of nonmagnetic or
antiferromagnetic Fe alloys with Ga and especially As outdiffused
from the bulk~\cite{Filipe,Gester,Chambers}.
\par

There have been several attempts to suppress the alloy formation:
sulfuric passivation of GaAs surface~\cite{Anderson}, insertion of
thin interlayer such as Er layer between Fe film and
GaAs~\cite{Schultz}. Those methods have reduced the outdiffusion
and alloy formation of Ga and As, but the segregated As atoms have
been still observed at the surface~\cite{Schultz}. Y. Chye $et$
$al$~\cite{Chye} grow the Fe film on GaAs(100) at a low
temperature ($\sim$ 120 K) and find improved magnetic properties
of the Fe film such as the improved squareness. The alloy
formation seems suppressed. However, structural and chemical
analysis to confirm the assumed growth behavior of Fe/GaAs(100) at
the low temperature has not been investigated. Only recently,
photoelectron spectroscopy has been performed for the Fe film
grown on GaAs(100) at 130 K and  suggested that the alloy
formation and outdiffusion of both Ga and As should be effectively
suppressed.\cite{Lee}
\par

In the present work, we study the growth and
thermal stability of the Fe films grown on GaAs(100) at 140 K by {\it in situ}
 x-ray reflectivity (XRR) study. The film has revealed the surface
width developing with the growth exponent $\beta\approx$ 0.50,
reflecting the restricted relaxation processes at low
temperature~\cite{Sarma1}. Most of all, we find the formation of
virtually pristine Fe film on GaAs(100) at 140 K.
Further, the Fe film shows thermal
stability against annealing up to 300 K, except the
interface region of 9\AA thickness where the alloy forms.
The confinement of the interface alloy
is ascribed to the inefficiency of bulk diffusion process
by which Ga and As diffuse into the Fe film.
\par

All the experiments are performed in an ultrahigh vacuum (UHV)
x-ray scattering chamber at 5C2 beam line of Pohang light source
(PLS). The base pressure of the chamber is below
$8\times10^{-10}$ Torr. The GaAs(100) sample is of square shape
with its lateral size 10 mm and thickness 0.6 mm. Clean GaAs(100)
is obtained by several cycles of 0.5 keV Ar$^+$ ion sputtering
with the incident angle of $45^{o}$ from the surface normal around
300 K and annealing up to 850 K. The sample temperature is
monitored by both an optical pyrometer and a K-type thermocouple
attached near the sample.
\par

In order to observe the evolution of the surface morphology upon Fe
deposition, specular XRR is measured.
For the experiment, UHV
chamber mounted on four-circle diffraction goniometer (2+2 mode)
is employed. The incident x-ray is vertically focused by a
mirror and monochromatized to a wavelength $\lambda = 1.239 \AA$
by a double bounce Si(111) monochrometor.
\par

A commercial electron beam evaporator (EFM3, Omicron) is used to
evaporate Fe (purity of 99.99\%) at the rate of 0.026 \AA /sec.
During the deposition, the chamber pressure is maintained below
$1\times10^{-9}$ Torr and the substrate is held at 140 K. The
deposition of Fe and the XRR measurement is alternated, and the
referred thickness $t$ of the Fe film is the accumulated amount of
the deposited Fe. Since the Fe film does not follow layer-by-layer
growth mode at 140 K, $t$ is a nominal value.
\par

Fig. 1 shows the specular XRR curves as a function of the
out-of-plane momentum transfer, $q_z$. In each curve, the
contribution from diffuse scattering was measured and subtracted.
For the clean GaAs(100), reflected intensity is observed  for very
large $q_z$, which indicates small surface width, $W_S$. After
curve fitting according to Parratt's formula,~\cite{Parratt} $W_S$
is found to be  2.01$\pm$0.40 $\AA$.
\par

As the Fe film becomes thick, however, the slope of the specular
XRR curve becomes steep, implying that the surface becomes rough.
Thus, the XRR curves show very weak modulation because the
constructive interference of the x-rays reflected from the rough
surface (Vaccum/Fe) and those reflected from the rough interface
(Fe/GaAs) hardly occurs. To assess the roughness of the film
quantitatively, we again applied the Parratt's formula to fit the
observed XRR curves. The fitting parameters include the effective
electron density in the film, $\rho$, the thickness of the film,
$t$, the width of the surface, $W_S$, and that of the interface,
$W_I$. $W_S$ ($W_I$) is defined as the root mean square of the
height fluctuations at the surface (interface)~\cite{Sinha}.
\par

Each solid line in Fig. 1 represents the best fit  of the
experimental specular XRR curve obtained for the Fe film grown on
GaAs(100) at 140 K. The model is composed of the homogeneous
single Fe film on GaAs(100) and well reproduces the experimental
XRR curves. (Fig. 1) The interface alloy and the segregated
As-derived layer need not to be invoked for the fit, suggesting
that no significant chemical reactions of the Fe with the Ga and
As occurs. The best-fit model is also consistent with the
observation of the recent photoelectron spectroscopy that finds
effective suppression of the outdiffusion of both Ga and As, and
negligible interface alloying for the Fe film grown on GaAs(100)
at 130 K~\cite{Lee}.
\par

In Fig. 2, summarized are the $W_S$ and $W_I$ as a function of
$t$. The linear dependence of the $W_S$ and $W_I$ on $t$ in a
double logarithmic scale implies a power law behavior,
$W_{S,I}\sim t^\beta_{S,I}$, where $\beta_S$ and $\beta_I$ are
0.51$\pm$0.04 and 0.37$\pm$0.03, respectively. Different $\beta$
for each interface indicates that the surface roughness does not
result from the propagation of the interface roughness or there is
working different growth mechanism for the roughening process on
each interface.
\par

A simple model, random deposition without lateral
relaxation~\cite{Barabasi} gives rise to the $\beta\approx$ 0.50
for the surface roughness. Besides, several models and
simulations~\cite{Sarma1,Zhang,Sarma2} assuming restrictive
lateral relaxation or lateral relaxation without interlayer
diffusion of adatoms also predict the $\beta\approx$ 0.50. The
observed $\beta_S$ insinuates that many diffusion processes for
the relaxation of the adatoms can be hardly activated at 140 K. It
is noteworthy that such a large $\beta\approx$0.50 has also been
observed during homoepitaxial growth of Ag on Ag(111) below 300 K
due to the restricted interlayer diffusion~\cite{Elliott}.
\par

Due to the restricted relaxation of the Fe adatoms at 140 K, the
Fe film is expected to contain non-negligible amount of vacancies
in it. The formation of vacancies or voids in the growing film has
been reported in the metal homoepitaxy at low
temperature~\cite{Botez1,Botez2,Casperson,Montalenti}. Then, the
roughening of the interface may be ascribed to those vacancies and
their island formation. The difference between the mobility of the
adatom and that of the vacancy seems to be the origin of the
different roughening kinetics observed at the surface and the
interface.
\par

We examine the thermal stability of the Fe film in the ambient
temperature that is usually requested for the device application.
After depositing the Fe film of 23 $\AA$ at 140 K, the sample is
annealed up to 300 K, and then its structure is investigated by
XRR. In Fig. 3(a), the overall intensity of the XRR is enhanced by
the annealing, and the modulation of the intensity with respect to
$q_z$ improves compared with that for the as-grown sample at 140
K, {\it e.g.} that for 19 \AA film in Fig.1, indicating that the
roughness of surface and that of interface are significantly
reduced by the smoothing effects at 300 K. Even after two more
hours of annealing of the sample, the XRR curve does not show any
change (Fig. 3(a)), which should be a sign of the steady state of
the system.
\par

In Fig. 3(b), shown is the best-fit XRR curve based on the
previously adopted model, the homogeneous single Fe layer on
GaAs(100), along with the experimental curve. The theoretical
curve does not properly reproduces the experimental one,
especially in the high $q_z$ region. Since the interface alloy is
expected to form at room temperature,
we try to fit the experimental XRR curve,
now including the interface alloy inbetween the Fe film
and the GaAs substrate. In Fig. 3(b), the best-fit curve from the
modified model well reproduces the full experimental data.
\par

The best-fit parameters are as follows; the surface roughness
$W_S$ (Vacuum/Fe), an interface roughness $W_{Fe/Interlayer}$ and
the other interface roughness $W_{Interlayer/GaAs}$ are
4.58$\pm$0.52 $\AA$, 2.89$\pm$0.43 $\AA$, and 3.35$\pm$0.47 $\AA$,
respectively. The interface alloy is 8.64$\pm$0.76 $\AA$ thick.
We assures the reduced roughnesses in
comparison with those for the as-deposited
 Fe films on GaAs(100) at 140 K,
{\it e.g.} that for 19\AA thick film in Fig. 2.
\par

Fig. 3(c) shows the variation of the effective electron density
$\rho$ giving the best-fit curve. $\rho$ shows the contribution
from the interface alloy of thickness 8.64$\pm$0.76 $\AA$
between Fe film and GaAs. The mean electron density in this layer
is 1.74$\pm$0.05 electrons/$\AA^{3}$ that is intermediate between
that of Fe (2.14 electrons/$\AA^{3}$) and that of GaAs (1.33
electrons/$\AA^{3}$), reflecting the intermixing of Fe with Ga
and/or As.
\par

The limited thickness and the confinement near the interface of
the alloy layer should derive from the inefficiency of the bulk
diffusion of the Ga and As from the substrate to the overlying
film. As a result, the pristine Fe film over this intermixed layer
is conserved or kinetically stabilized even at room temperature.
In line with the present model, no compound formed by the
segregated Ga and As is observed by photoelectron spectroscopy,
either~\cite{Lee}.
\par

{\it Summary - } We investigate the growth of the Fe films on GaAs(100) at
140 K, by UHV $in-situ$ x-ray reflectivity. At 140 K, rough Fe film
forms on GaAs(100) due to restrictive surface relaxation.
The chemical reaction at the interface and surface segregation of As
seems virtually suppressed during the growth at 140 K.
After warming up the sample to room temperature,
intermixed layer of $\sim 9\AA$ forms at the
interface, but the system is in a steady state, showing no further
development of the interface alloy, keeping the overlying Fe film in its pristine state.
\par

\newpage

\begin{figure}
\caption{Specular x-ray reflectivity (symbols) as a function of
the out-of-plane momentum transfer, $\it q_z$ for increasing Fe
film thickness, $t$ at 140 K. The solid line is the theoretical
prediction by the Parratt's formalism. The curves are vertically
shifted with respect to each other for clarity.}

\caption{The surface roughness, $W_S$ and interface roughness
,$W_I$ as a function of the deposited Fe film thickness, $t$ at
140 K. The solid lines shows the power law fitting results of
$W_S$ and $W_I$ with the scaling relation, $W\sim t^\beta$.}

\caption{(a) Specular XRR curves after annealing of 23 $\AA$ Fe
film at 300 K for 1 (open circle) and 2 hours (solid triangle),
respectively. (b) The solid (dashed) line in the upper (lower)
graph is the best-fit curve according to the model of homogeneous
single Fe layer (an Fe layer and then an alloy layer) on
GaAs(100). The two graphs are vertically shifted for the clear
presentation. (c) the best-fit, electron density profiles for the
double layer model.}

\end{figure}

\newpage

\begin{figure}
\includegraphics[width=1\textwidth]{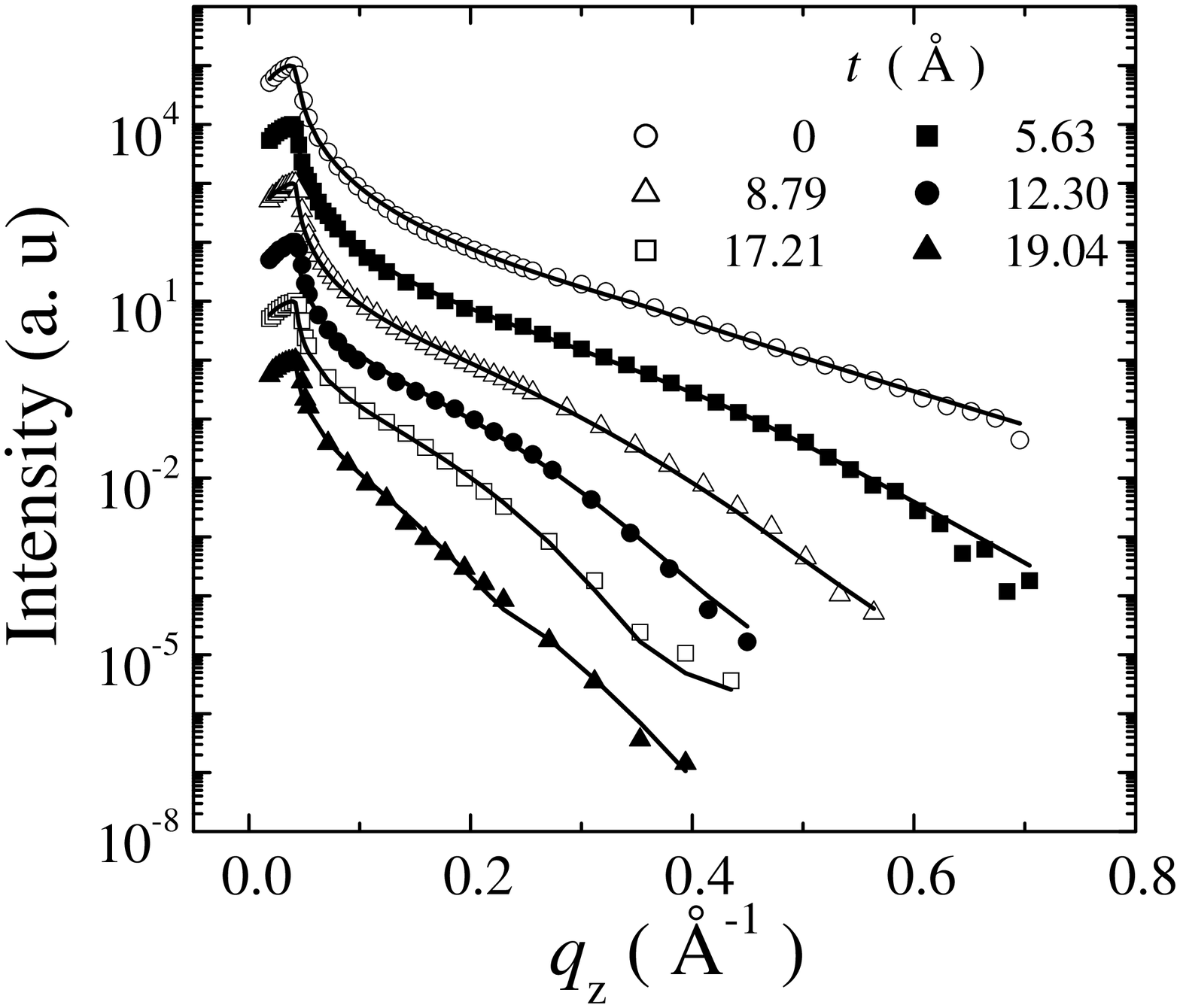}
{\textbf{Kim \textit{et al.} Fig. 1}}
\end{figure}

\begin{figure}
\includegraphics[width=1\textwidth]{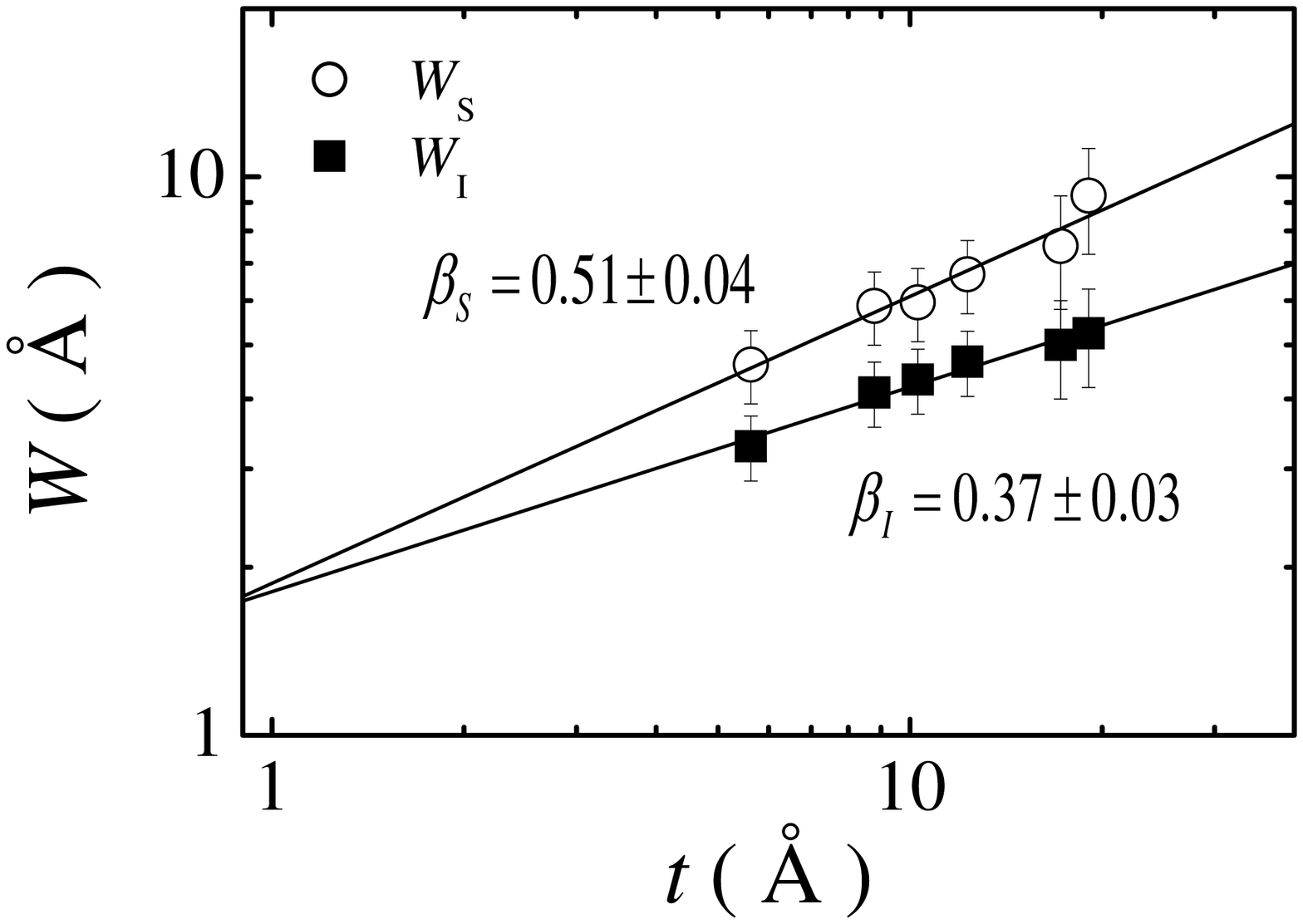}
{\textbf{Kim \textit{et al.} Fig. 2}}
\end{figure}

\begin{figure}
\includegraphics[width=1\textwidth]{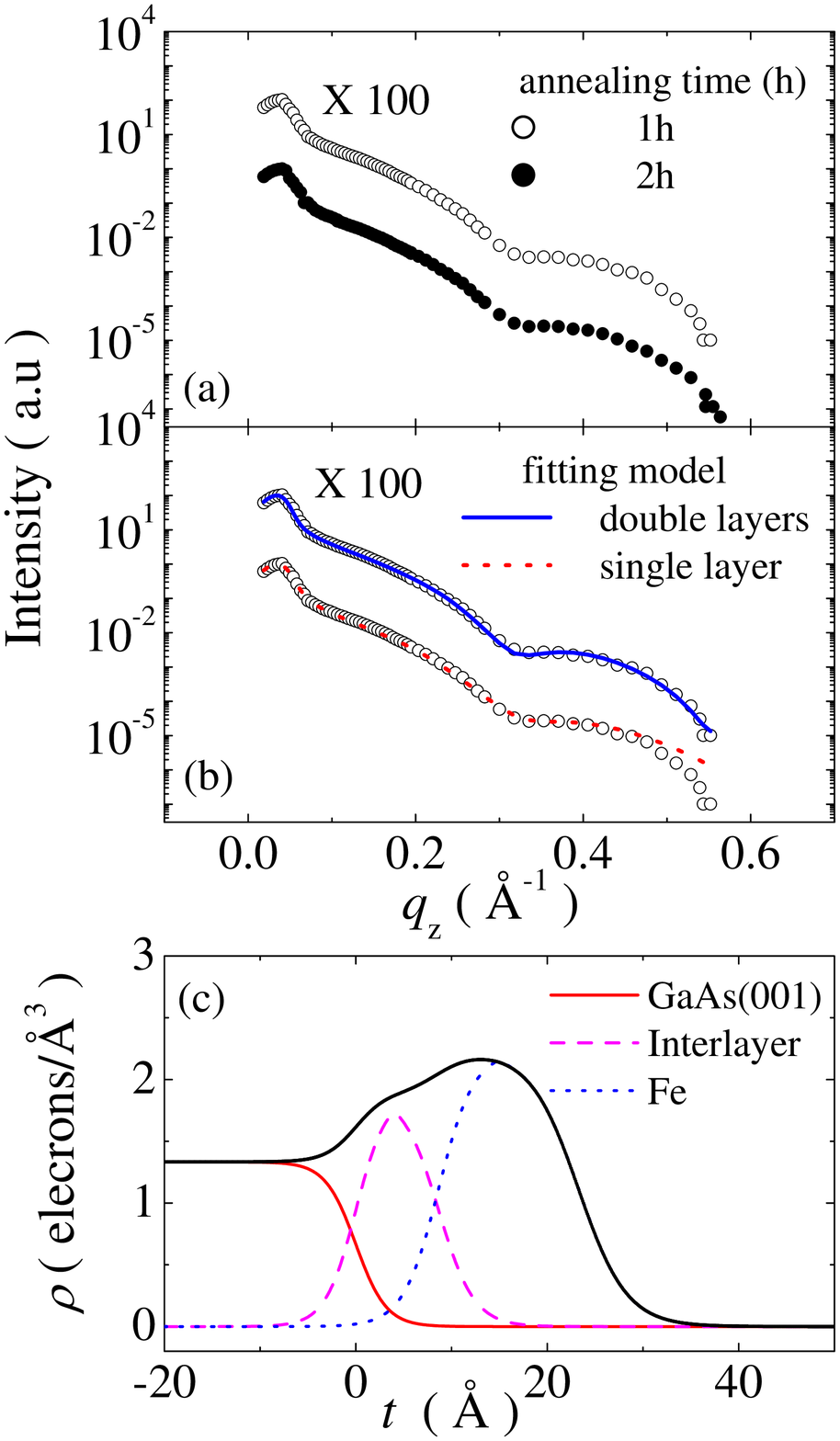}
{\textbf{Kim \textit{et al.} Fig. 3}}
\end{figure}

\end{document}